# Structural and Magnetic Characterization of $Ni_2MnBO_5$ ludwigite

Moshkina Evgeniya[a, b], Sofronova Svetlana[a], Veligzhanin Alexey[c], Molokeev Maxim[a], Nazarenko Ilya[b], Eremin Evgeniy[a], Bezmaternykh Leonard[a]

[a] L.V. Kirensky Institute of Physics SB RAS, 660036 Krasnoyarsk, Russia
[b] Siberian State Aerospace University named after Academician M.F. Reshetnev, Krasnoyarsk, Russia
[c] National Research Centre "Kurchatov Institute", Moscow, Russia

* e-mail: ekoles@iph.krasn.ru

**Abstract**

Single crystals of ludwigite $Ni_2MnBO_5$ were synthesized by flux growth technique. The detailed structural and magnetic characterizations of the synthesized samples have been carried out. The cations composition of the studied crystal was determined using X-ray diffraction and EXAFS technique, the resulting composition is differ from the content of the initial $Mn_2O_3$ – CuO components of flux. Magnetic susceptibility measurements and the calculations of the exchange integrals in frameworks of indirect coupling model revealed strong antiferromagnetic interactions and appearance of magnetic ordering phase at the temperature $T$=85 K. The hypothesis of the existence of several magnetic subsystems was supposed.

**Introduction**

Oxyborates $Me1^{3+}Me2_2^{2+}BO_5$ (Me1, Me2 – transition metals) with a ludwigite structure are the members of a large class of compounds have very interesting features, such as charge ordering, existence of several magnetic subsystems ordered at different temperatures and in perpendicular directions, reversal magnetization, spin-glass state [1-5]. Due to these extraordinary properties oxyborates with a ludwigite structure are promising materials for possible applications. Moreover, these compounds are good model objects due to the complex quasi-two-dimensional structure including low-dimensional elements – linear chains, zig-zag walls and three-leg ladders [1-3] and large number of magnetic ions with different valence states, including the heterovalent cations of one transition metal, per unit cell. The ludwigites structure may contain not only trivalent and bivalent cations of transition metals but it can contain tetravalent cations substituting $2M^{3+}=M^{2+}+M^{4+}$. The magnetic properties of Mn-Ni ludwigites are very sensitive to the composition $x$, may be, because of changing of valence states of cations.The presence of tetravalent cations can significantly influence the structural, magnetic and electric properties of compound. So, first of all it is necessary to establish the exact composition and valence states of cations of these compounds for carrying of detail studies and determination of micromechanisms which take place in ludwigites.

Recently a new Mn-heterovalent Ni-containing ludwigites $Mn_{3-x}Ni_xBO_5$ were synthesized. First magnetic characterization was made for compounds with $x$=0.5, 1.5, 1.8 [5, 6]. It was found, that ludwigite $Mn_{3-x}Ni_xBO_5$($x$=1.5 – in agreement with the initial components ratio in the flux) undergoes a ferrimagnetic phase transition at $T_c$=85 K, temperature-field dependencies of magnetization demonstrate strong anisotropy and changing of "easy" axis direction in dependence of the magnitude of magnetic field. So, the goal of the present work is detailed investigation of structural and magnetic characteristics of the ludwigite $Mn_{3-x}Ni_xBO_5$($x$=1.5 – in agreement with the initial components ratio in the flux) using experimental and theoreticalmethods: refinement the composition $x$ and the valence states of transition metals ions, determination of the exchange constants and the probable magnetic structure.

## I. Crystal structure
### a. Crystal growth and X-ray diffraction

Single crystals $Mn_{3-x}Ni_xBO_5$ ($x=1.5$ – in agreement with the initial components ratio in the flux) with ludwigite structure were synthesized by flux method, using the fluxes based on bismuth trimolybdate with addition of sodium carbonate. Thedetailsofcrystallizationconditionsaredescribedin[6].The grown single crystals in the form of black orthogonal prisms have a length of 6mm and a transverse size of about 0.3mm.

The intensity pattern was collected from single crystal at 25 ºC using the SMART APEX II X-ray single crystal diffractometer (Bruker AXS) equipped with a CCD-detector, graphite monochromator and Mo-Kα radiation source. The absorption corrections were applied using the SADABS program. The structures were solved by the direct methods using package SHELXS and refined in the anisotropic approach for all atoms using the SHELXL program [7]. The DIAMOND program is used for the crystal structure plotting [8].

Crystal structure of $Ni_{1.5}Mn_{1.5}BO_5$ was solved in *Pbam* space group using single crystal experiment earlier [6]. Asymmetric part of unit cell contains one $BO_3$ triangle and four (Ni/Mn)$O_6$octahedra (Fig.1). Each octahedron is linked with other octahedra by edges and corners, and linked with $BO_3$ triangles. Each $BO_3$ is linked with nine (Ni/Mn)$O_6$ octahedra by corners, so that crystal structure forms 3D net. This model was not restrained, and Ni/Mn occupation ratios were refined for all four Mn/Ni sites. Refinement showed that final chemical formula $Ni_{1.6(1)}Mn_{1.4(1)}BO_5$ is close to proposed, and the structure proved to be isostructural to $Ni_{2.5}Mn_{0.5}BO_5$ [9].

The distortions of all (Ni/Mn)$O_6$octahedra in model (A) were calculated by formula following formula [10]: $D = \frac{1}{n}\sum_{i=1}^{n}\frac{l_i - l_{av}}{l_{av}}$, where $l_i$ is the distance from the central atom to the $i^{th}$ coordinating atom, $l_{av}$ is the average bond length. It was found that $D_{Ni1/Mn1} = 0.0264$, $D_{Ni3/Mn3} = 0.0261$ have higher values in comparison with $D_{Ni2/Mn2} = 0.0144$ and $D_{Ni4/Mn4} = 0.0047$. These values don't correlate with Ni/Mn concentration ratio; however, they correlate with local symmetry of octahedra. Two octahedra (Ni1/Mn1)$O_6$ and (Ni3/Mn3)$O_6$ have low local symmetry *m*, whereas (Ni2/Mn2)$O_6$ and (Ni4/Mn4)$O_6$ have higher local symmetry *2/m*. Therefore (Ni1/Mn1)$O_6$ and (Ni3/Mn3)$O_6$ octahedra have higher possibility to distort and they have high *D* values.

It should be noted that the average bond lengths of (Ni/Mn)$O_6$octahedra: $d$(Mn1/Ni1-O) = 2.065 Å, $d$(Mn2/Ni2-O) = 2.066 Å, $d$(Mn3/Ni3-O) = 2.045 Å, $d$(Mn1/Ni1-O) = 2.078 Å are all in narrow range, and $d$(Mn3/Ni3-O) is the smallest bond from them. This is in accordance with occupation of Mn3 site by $Mn^{4+}$ ion because its ion radii IR($Mn^{4+}$, CN = 6) = 0.53 Å is smaller than IR ($Mn^{2+}$, CN = 6, LS-HS) = 0.67-0.83 Å, IR ($Mn^{3+}$, CN = 6, LS-HS) = 0.58-0.645 Å, and IR ($Ni^{2+}$, CN = 6) = 0.69 Å [11].But in practice the establishment of ions valence states in framework of X-ray diffraction experiment only is not possible, to this reason the EXAFS method was used.

### b. EXAFS

The X-ray absorption spectra were measured at experimental station "Structural material science" of Kurchatov source of synchrotron radiation [12]. The electron energy in a storage ring was 2.5 GeV, the current was about 70 mA. The beam size of the SR (synchrotron radiation) on the sample was 1 mm×2 mm. The EXAFS spectra were measured by the transmission technique using air ionization chambers. The point measurements of the absorption spectra were carried out in the range -170 eV – +800 eV relatively the energy of *K*-edge of absorption of the base elements – manganese (6539 eV) and nickel (7709 eV). This range was divided into three segments to the measurement time – before the edge region (-170 eV – -20 eV), near-edge region (-20 eV – 80 eV) and the EXAFS oscillation region (80 eV – 800 eV behind the absorption jump). In the first region the spectrum was measured with the step 10 eV, at near-edge region the step was ~0.5 eV, in the third one the scanning was performed equidistantly with the step 0.05 Å$^{-1}$ of the photoelectron impulses.

Each spectrum was measured about 20 minutes. The measurements of the every sample spectrum were held 2-3 times and averaged. Processing and analyzing of the results were performed using the program complex IFEFFIT [13-14], version 1.2.11c.

The analysis of manganese valence state was performed by the "fingerprint" method. To this purpose, the comparison of the position of *K*-edge of absorption of manganese and nickel of the studied sample and the set of characterized and well-studied standards - $MnB_2O_4$ ($Mn^{2+}$), $Mn_{1-x}Fe_xMoO_4$ ($Mn^{2+}$), $Mn_2O_4$ ($Mn^{3+}$) и $MnO_2$ ($Mn^{4+}$), NiO ($Ni^{2+}$), metallic nickel – was performed. To this comparison, the shapes of *K*-edge of absorption of the samples and the first derivative of the absorption spectra were visually analyzed to determine the energy positions of the inflection points.

As one can see (Fig. 2, 3) the position of *K*-edge of absorption of nickel of the studied sample matches to the position of absorption edge of NiO. The position of manganese absorption edge of the studied sample matches to the position of $Mn_2O_3$ absorption edge. However, the width and the shape of the lines of absorption for manganese are different. It could be explained by the difference of the crystalline environment of the studied sample and standard and by presence of the small amount of the $Mn^{4+}$, predicted by X-ray diffraction data. The resultant composition was determined using the jump of the K-edge absorption and it is Mn:Ni=1:2, that quite differ from the content of the initial components $Mn_2O_3$ – CuO (Mn:Cu=1:1) and X-ray diffraction data.

The parameters of the local structure of the material around the absorbing atom were determined by fitting of the calculated spectrum to the experimental spectrum EXAFS. The calculations were carried out in the approximation of the single-electron scattering. Number of coordination spheres was selected for each absorbing atom separately. Final model includes two coordination spheres for nickel atom and five coordination spheres for manganese atom. Such modeling allows establishing the distances $R_j$ between the absorbing atom and nearest neighbors, rms change in bond lengths $\sigma_j^2$ for couples of atoms, accounting for thermal vibrations and static disorder of the local environment. Coordination numbers $N_j$ of the nearest neighbors relatively the central metal atom were fixed in accordance with the structural model of studied samples. The value $E_0$ of the power shift of the photoelectron impulses origin relatively the *K*-edge absorption position and the damping factor $S_0^2$ of the signal amplitude were included to the refinement.

The local environment of nickel is characterized by quite low value of parameter $S_0^2$. It could be explained by some distortions of octahedral environment. However, the attempts to describe this distortion by the adding of extra inequivalent distances Ni-O do not lead to improvements of agreement of calculated and experimental curves.

The results of the fitting of Fourier transformations and oscillating part of the EXAFS spectra are shown in Fig.4, 5.The data obtained on the local environment of the ions Mn and Ni are presented in Table 1 and 2. As one can see, the octahedron around the manganese ion is elongated, that is in agreement with the supposed manganese ion valence +3, because such octahedral distortion is typical for Jahn-Teller ion, which is a $Mn^{3+}$. The octahedron around the nickel ion, apparently, has a small distortion; because the adding of extra inequivalent distances Ni-O do not lead to improvements of agreement of calculated and experimental curves. In metallic environment of manganese ion it is possible to distinguish a one short bond – 2.83 Å, two long bonds – 3.21 Å and six bonds 3.08 Å. In metallic environment of nickel ion it is possible to distinguish two long bonds – 3.21 Å and six short bonds 3.07 Å. From the comparison of the EXAFS and X-ray diffraction data it is obvious, that the position Me3 is occupied by the $Mn^{3+}$. The oxygen octahedron around the Me3 ion is elongated and in metallic environment it is possible to distinguish one short – 2.77 Å, two long 3.34 Å and six middle bonds from 2.99 Å to 3.11 Å.All the other positions, obviously, are occupied by the bivalent nickel ions. The oxygen octahedrons around Me2 and Me4 have slight distortions that are in agreement with EXAFS data, the octahedron around Me1 is compressed and this is not typical for trivalent manganese. Metallic environment of the nickel ion in different positions is different. In two positions it is possible to distinguish two long bonds – 3.34 Å (3.21 Å – EXAFS data), other bonds mainly are in the range 2.99 Å-3.11 Å in agreement with EXAFS data – 3.07 Å.

Distribution of ions with a different valence states on inequivalent crystallographic positions is not randomin the ludwigite structure. In the most ludwigite compounds the trivalent ions are occupy the position Me3 [15], while the others positions are occupied by bivalent ions. It was supposed, that the manganese ions are predominantly occupy the Me3 position, and the others positions are occupied by nickel, because nickel in the studied crystal is presented only by bivalent ions, manganese – by trivalent ions. Such distribution is in agreement with refined composition of compound $Ni_2MnBO_5$.

## II. Magnetic properties
### a. Field-temperature dependence of magnetic susceptibility

Magnetic measurements of single crystal $Ni_2MnBO_5$were performed using the physical properties measurements system PPMS-9 (Quantum Design) at temperature range $T=3 \div 300$ K and magnetic fields up to 80 kOe.

Temperature dependence of magnetic susceptibility $\chi=M/B$of ludwigite $Ni_2MnBO_5$is shown in Fig. 6. This dependence was obtained upon cooling the sample in the magnetic field $H=1$ kOe (FC) at the temperature range $T=3 \div 300$ K orthogonal ($H \perp c$) to the c axis.

As one can see temperature dependence of magnetic susceptibility is follow to Curie-Weiss law down to $T_c=85$ K. The temperature $T_c=85$ K matches the temperature of ferrimagnetic ordering [6]. Below this temperature another anomaly is observed: the temperature dependence of susceptibility demonstrates another inflection point at $T \approx 71$ K. It could be supposed, that the presence of two inflection points at this temperature dependence may indicate the existence of several loosely coupled magnetic subsystems being ordered separately at different temperatures.The specific heat studies are planned to prove this hypothesis and to study of thermodynamics characteristics of the crystal in details. The presence of the obvious "stairs" at the field dependencies of magnetization at low temperatures ($T=3$ K, 20 K) may also indicate the existence of several loosely coupled magnetic subsystems (Fig. 7).

The fitting of temperature dependence of magnetic susceptibility of paramagnetic phase was performed using modified Curie-Weiss law ($T=200 \div 300$ K) [16]:

$$\chi = \chi_0 + \frac{C}{T-\theta} \quad (1)$$

The parameters of this dependence were determined. Temperature-independent term, which includes diamagnetic contribution of completely filled electron shells and Van-Fleck paramagnetism, is a positive $\chi_0 \approx 0.22 \cdot 10^{-4}$ emu/mol (diamagnetic contribution to $\chi_0$ was calculated by summing the diamagnetic Pascal constants of each ion and it is $\chi_D = -0.81 \cdot 10^{-4}$ emu/mol [17]). Negative sign of Weiss temperature $\theta \approx -157$ K indicates the strong antiferromagnetic interactions in crystal. Also, as a result of approximation, the value of Curie constant of studied ludwigite was obtained, and it is $C \approx 5.06$ emu·K/mol. Using this parameters, effective magnetic moment of the formula unit was estimated via relation $\mu^2_{eff} = 8C$, and it is $\mu_{eff} \approx 6.32 \mu_B$. Effective magnetic moment also was calculated theoretically via $\mu_{eff} = \sum(Ng^2S(S+1)\mu_B^2)^{1/2}$ (only spin component of the effective magnetic moment was taken into account), where $N$ is composition $x$, $g$ is $g$-factor of ions $Mn^{3+}$($g \approx 2$ [18]) and $Ni^{2+}$($g=2.08$ [19]) in octahedral coordination, obtained in other works. Calculated value of effective magnetic moment $\mu_{eff2} \approx 6.36 \mu_B$ agree with the value of experimentally estimated effective magnetic moment $\mu_{eff} \approx 6.32 \mu_B$ within the error determination of Curie constant C and according with high-spin states $S(Ni^{2+})=1$, $S(Mn^{3+})=2$.

### b. Indirect coupling model

Nickel and manganese ions in bimetallic ludwigites can be presented in various proportions depending on the compound composition. $Ni_2MnBO_5$ was studied using indirect coupling model. The occupancy of the crystallographic positions in $Ni_2MnBO_5$ is as follow. Ni ions occupy positions 4g, 2d, and 2a; Mn ions occupy the only position – 4h, in accordance with EXAFS results.

In the ludwigites, there is a competition between three types of magnetic interactions: direct exchange, superexchange, and double exchange [20]. To analyze the magnetic structure and estimate the superexchange interactions in $Ni_2MnBO_5$ crystal we used a simple indirect coupling model [21, 22] based on the theory of the superexchange interaction of Anderson, and Zavadskii [22], and Eremin [23]. Within the indirect coupling model, the structure of the crystal can be characterized by the following integrals of the indirect exchange coupling with regard to occupations of individual cation orbitals and symmetries of the lattice of indirect couplings $J_{ij}$ where $i$ and $j$ are the numbers of nonequivalent crystallographic positions for magnetic ions.

The calculated exchange integrals for $Ni_2MnBO_5$ are presented in the Table 3. Here, $b$ and $c$ are the electron transfer parameters being squares of ligand-cation intermixing coefficients for the $\sigma$ and $\pi$ coupling, respectively (the values of these parameters are $b = 0.02$ and $c = 0.01$). $U(Ni^{2+}) = 2.7$ eV, and $U(Mn^{3+}) = 5$ eV are the cation-ligand excitation energy; $J^{in}(Ni^{2+}) = 2$ eV, and $J^{in}(Mn^{3+}) = 2.7$ eV are the integral of interatomic exchange interaction [21, 22].

One can see from the Table 3 that all Ni-Ni superexchange integrals are ferromagnetic. The superexchange $J_{4h-2d}$ is weak and ferromagnetic. Other superexchange integrals, which involve 4h position, are antiferromagnetic. The superexchanges between zigzag walls ($J_{4g-4h}$ and $J_{4g-2a}$) are also weak but antiferromagnetic.

The suggested magnetic structure of $Ni_2MnBO_5$ ludwigite is shown in the Fig. 8 with superexchange values denoted. The magnetic order is mainly formed by strong AFM and FM interactions within the zigzag walls since the interactions between the walls are weak and AFM. Zigzag walls are coupled with each other antiferromagnetically. Triad 4h-2d-4h order is shown in the figure. Within the triad, 2d and 4h ions are coupled antiferromagnetically. Along *c* axis triad ions are coupled ferromagnetically in spite of AFM interaction of Mn ions within the position 4h which is twice weaker than interaction with neighboring ions in the position 4g. That is why the interactions within the position 4h are frustrated as well as interactions between 4h and 2d ions in neighboring triads.

The suggested magnetic structure of $Ni_2MnBO_5$ is shown in Fig. 9. Since the indirect coupling model is able to estimate ions magnetic moments directions only relative to each other, in our model all magnetic moments are directed along *a* axis for easier perception. In present case, the magnetic cell coincides with the crystallographic one.

**Conclusions**

The detailed studying of structural and magnetic properties of ludwigite $Ni_2MnBO_5$ has been carried out. Structure of the studied sample, lattice parameters and bond lengths have been determined by single crystal X-ray diffraction. The assumption on the Mn and Ni ions crystallographic distribution has been done. EXAFS experiment has been performed to obtain more exact determination of composition and valence states of ions. It was found, that the actual ratio Mn:Ni=1:2 in crystal doesn't match the ratio of initial component of flux which was Mn:Ni=1:1. It was established the nickel in this compound is only bivalent, while the manganese is predominantly trivalent with possible little addition of tetravalent manganese. Obtained bonds lengths from the EXAFS experiment are agree with the X-ray diffraction data.

Using the results of X-ray diffraction and EXAFS experiments the field-temperature dependencies of ludwigite $Ni_2MnBO_5$ have been interpreted. The parameters of modified Curie-Weiss law (1) have been determined. Experimentally obtained value of effective magnetic moment is in quite well agreement with theoretically calculated value of the spin component of effective magnetic moment, which matches to high-spin states of Mn and Ni ions and to ratio Mn:Ni=1:2, determined by EXAFS technique. Temperature dependence of magnetic susceptibility has revealed anomalies at $T_c$=85 K and $T$=71 K. These anomalies and presence of "stairs" at the field dependencies of magnetization at low temperatures ($T$=3 K, 20 K) may indicate the existence of several loosely coupled magnetic subsystems being ordered separately at different temperatures.

The calculations of the exchange integrals in frameworks of indirect coupling model has revealed strong competing antiferromagnetic interactions which lead to frustrations in magnetic structure of studied compound. The model of magnetic structure has been suggested.

**Acknowledgements**

The authors are grateful to Professor N.V. Volkov and Professor A.N. Vasiliev for valuable discussions. This work was supported by Russian Foundation for Basic Research (grants No. 16-02-00055, 16-32-00318).

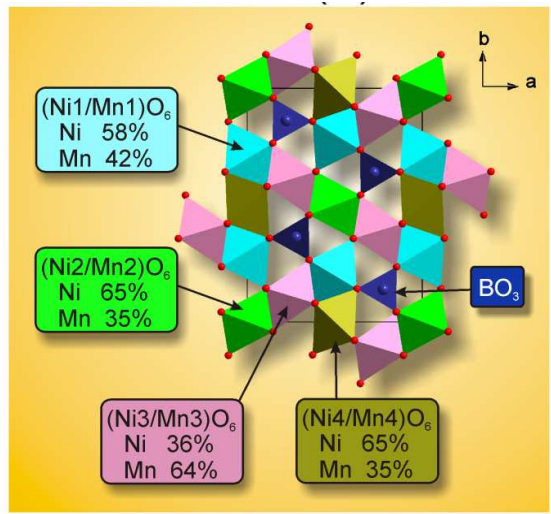

Fig.1. The model of crystal structure of $Ni_xMn_{3-x}BO_5$ with Ni/Mn occupations in each site.

Me1(4g), Me2 (2a), Me3(4h), Me4(2d)

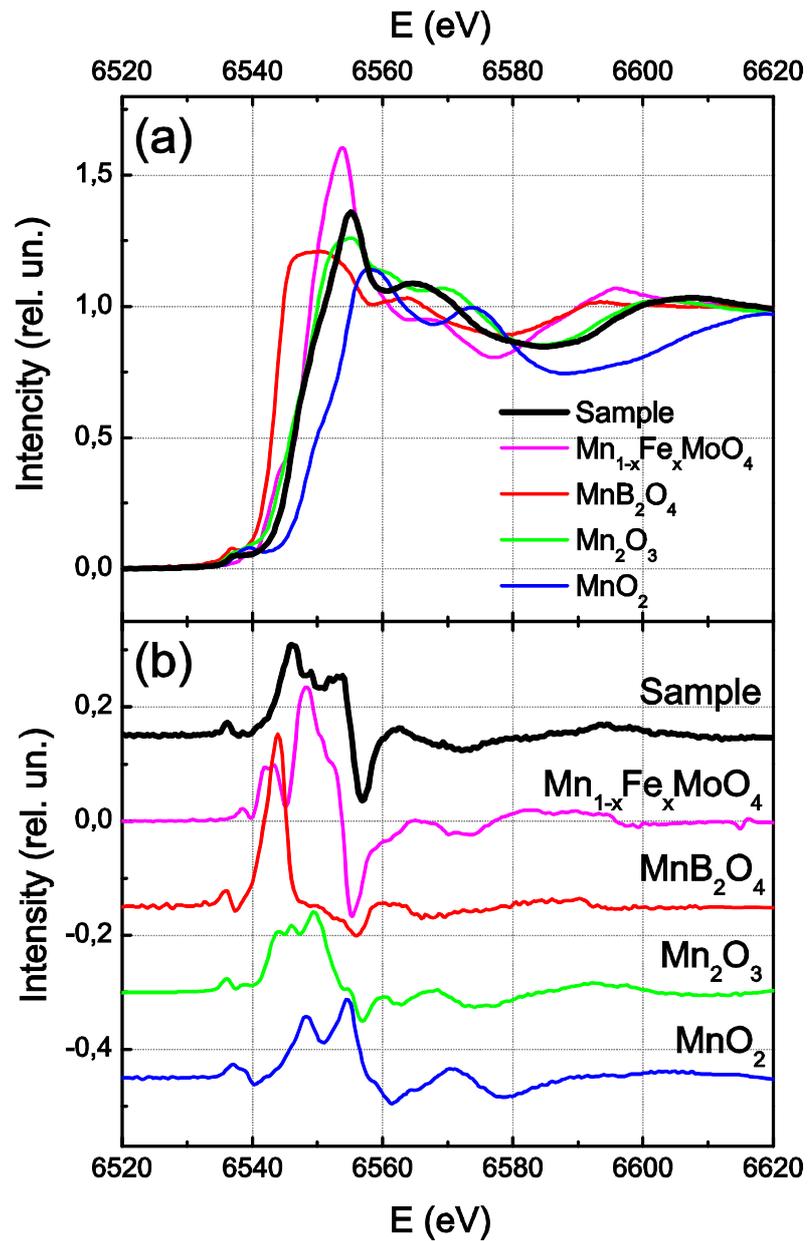

Fig. 2 K-edge of absorption (a) and the first derivative of the absorption spectra of the Mn (b).

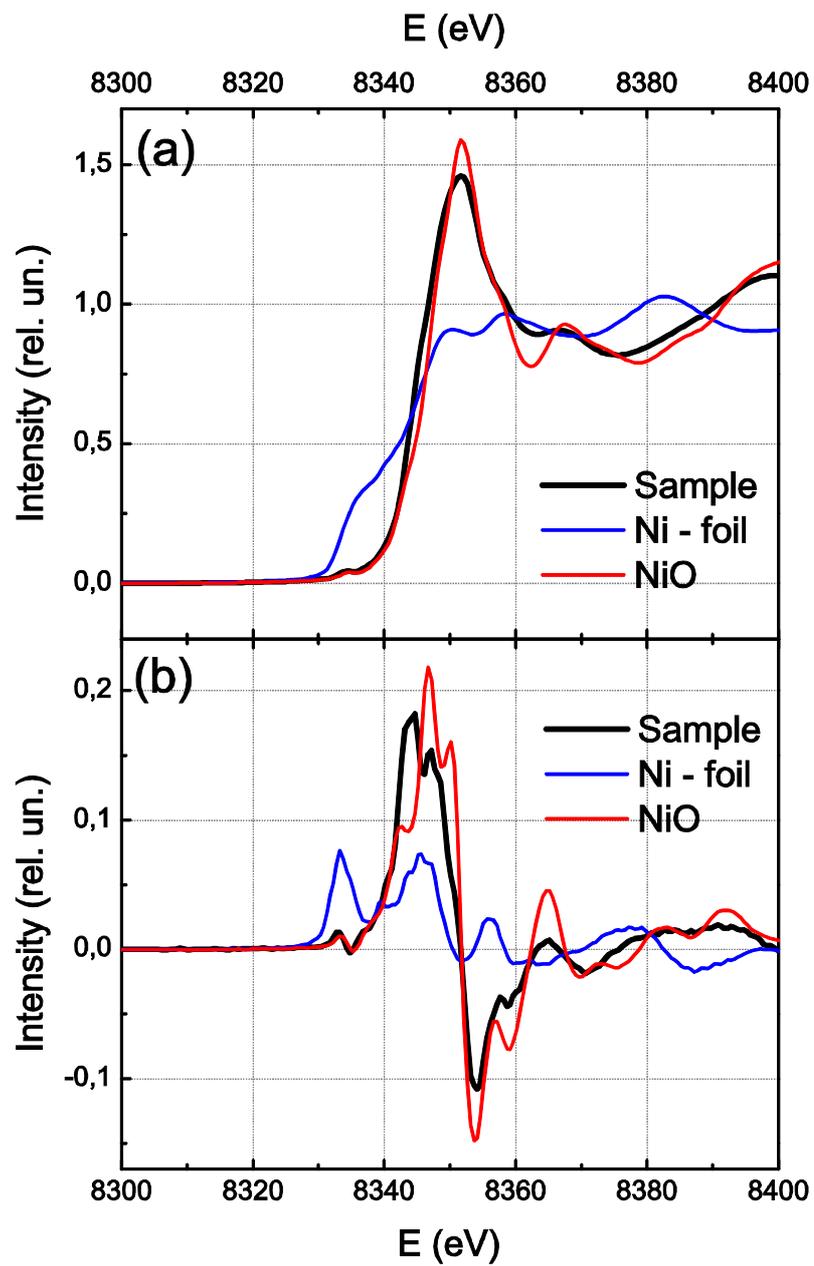

Fig. 3 K-edge of absorption (a) and the first derivative of the absorption spectra of the Ni (b).

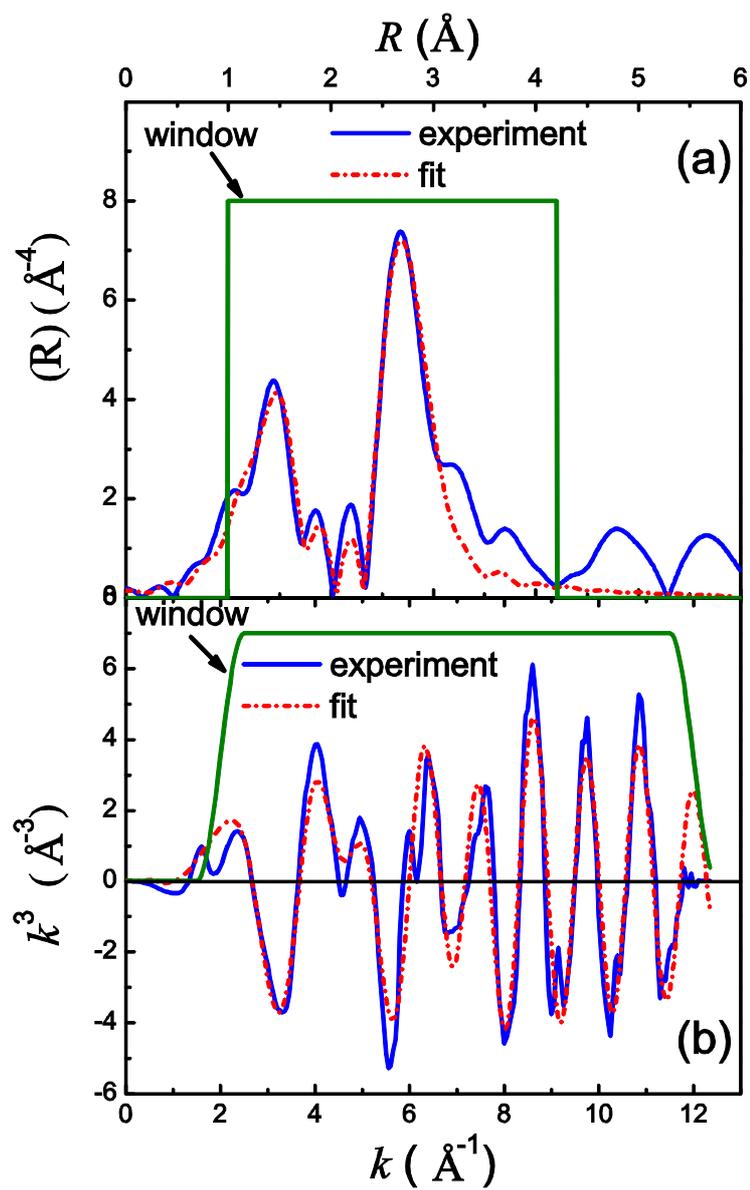

Fig. 4 Fourier transformations (a) and oscillating part (b) of the EXAFS spectra of manganese.

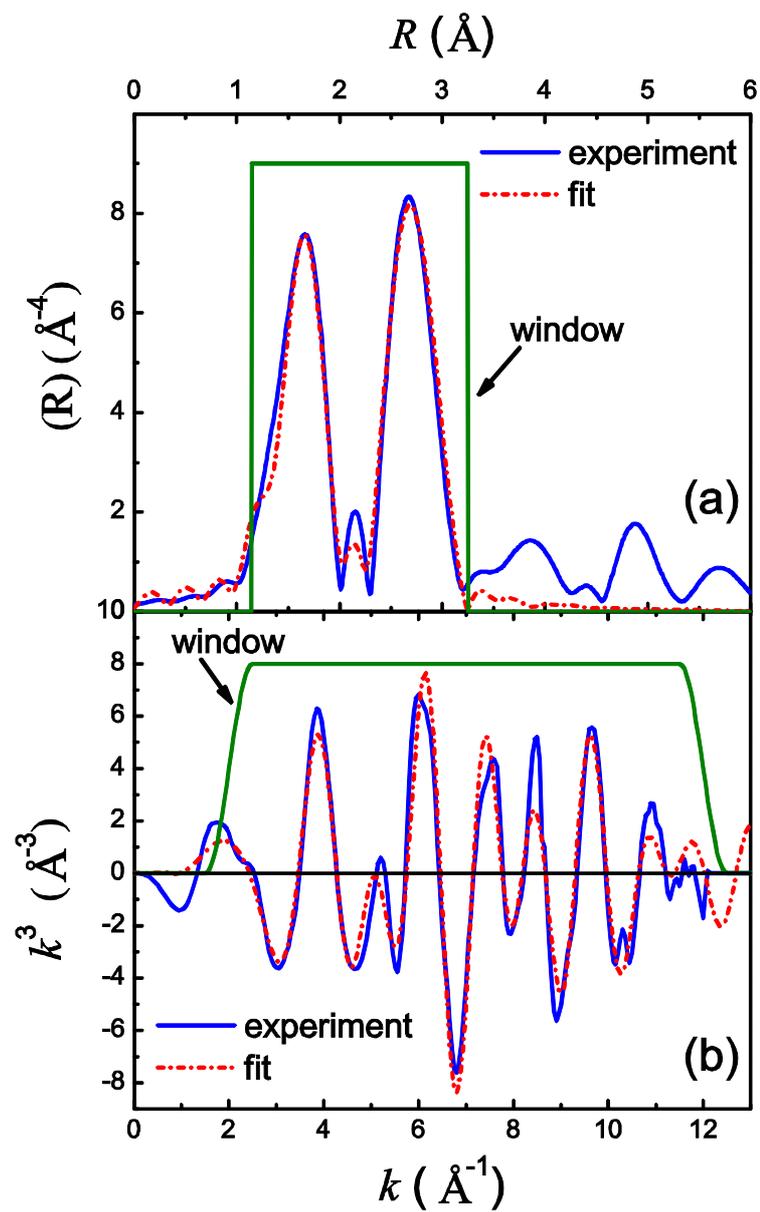

Fig. 5 Fourier transformations (a) and oscillating part (b) of the EXAFS spectra of nickel.

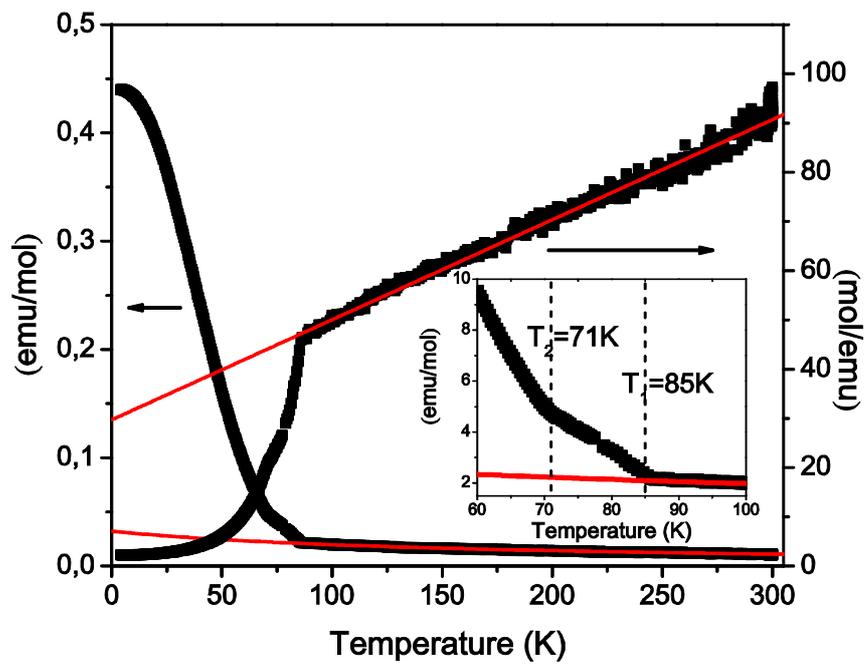

Fig. 6 Temperature dependence of magnetic susceptibility and reversal susceptibility, obtained upon cooling the sample at H=1 kOe, H⊥c (FC).

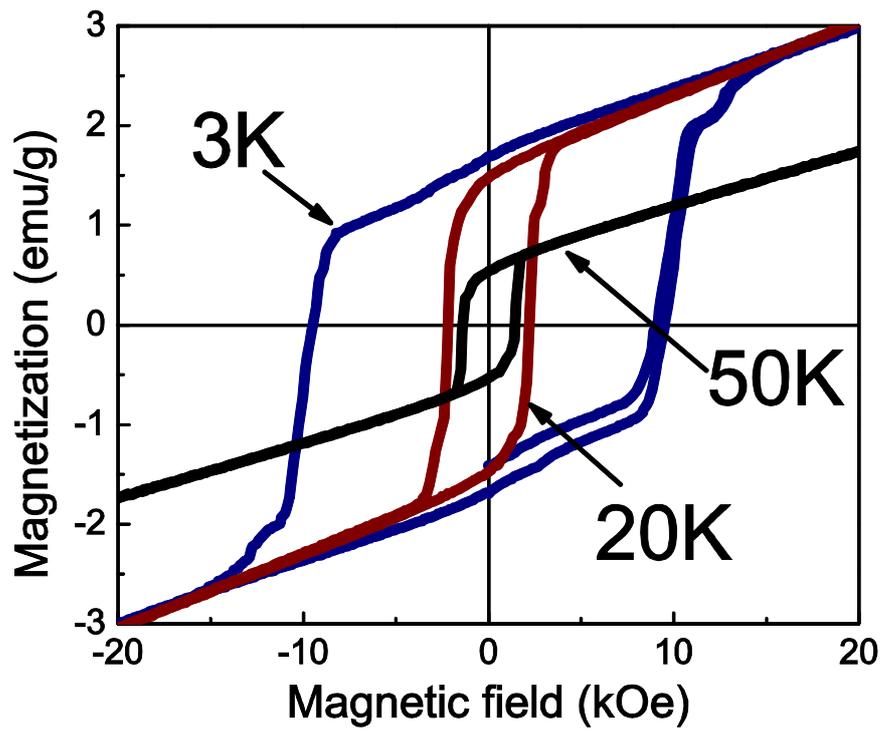

Fig. 7 Magnetic field dependencies of magnetization of ludwigite Ni$_2$MnBO$_5$, obtained at T=3 K, 20 K, 50 K (H⊥c).

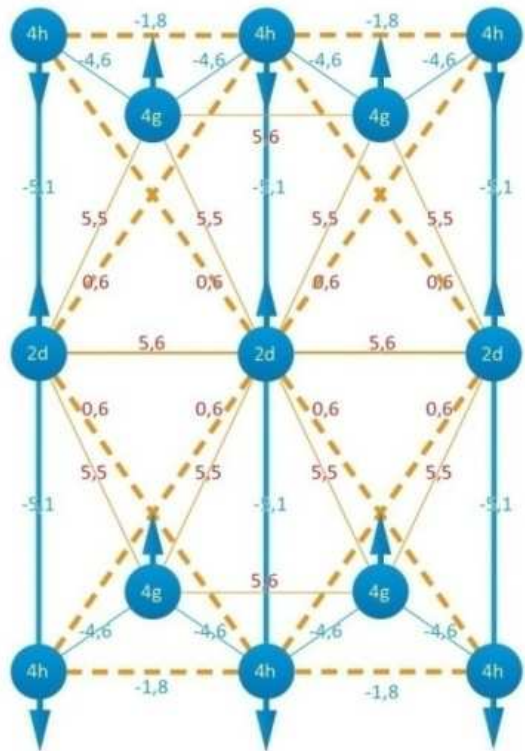

Fig. 8 Magnetic order of triads 4h-2d-4h and 4g-2d-4g for Ni$_2$MnBO$_5$. Axis c is directed horizontally from the left to the right. Dashed lines show frustrated interactions. Arrows show the direction of ions magnetic moments relative to each other.

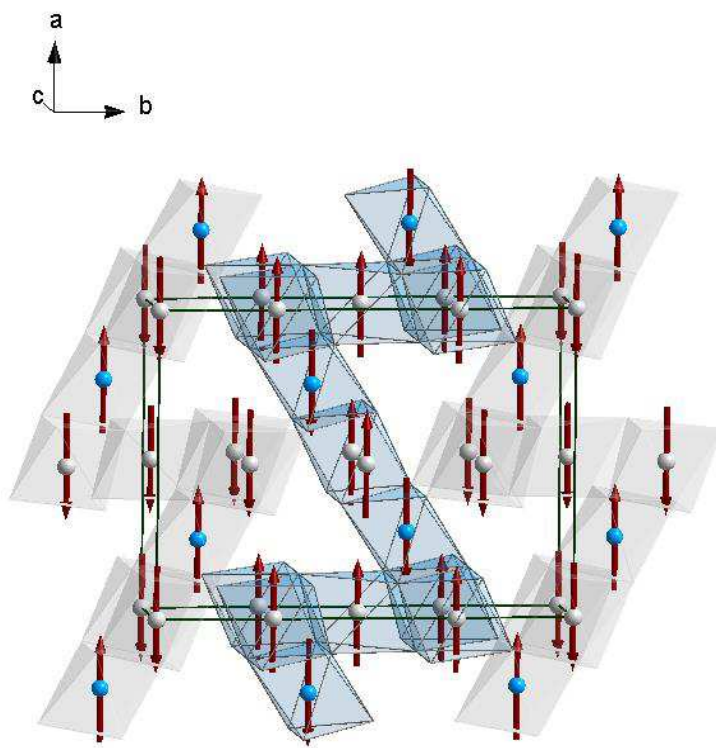

Fig. 9 Magnetic structure of Ni$_2$MnBO$_5$, obtained in the framework of indirect coupling model.

Table 1 The local environment of the manganese and nickel ions.

| Ion | Rf,% | k-range | R-range | $S_0^2$ | $E_0$,eV | Scattering way | N | R, A | $\sigma^2$, $A^2 \times 10^{-3}$ |
|---|---|---|---|---|---|---|---|---|---|
| Mn | 6.69 | 2 - 12 | 1.0 - 4.2 | 0.78+-0.18 | 1.75+-2.65 | Mn-O1 | 4 | 1.95(3) | 11+-4 |
|  |  |  |  |  |  | Mn-O2 | 2 | 2.17(7) |  |
|  |  |  |  |  |  | Mn-Me1 | 1 | 2.83(4) | 7+-1 |
|  |  |  |  |  |  | Mn-Me2 | 6 | 3.08(1) |  |
|  |  |  |  |  |  | Mn-Me3 | 2 | 3.35(3) |  |
| Ni | 1.47 | 2 - 12 | 1.15 - 3.25 | 0.61+-0.07 | 2.88+-1.09 | Ni-O | 6 | 2.088(9) | 5+-1 |
|  |  |  |  |  |  | Ni-Me | 6.0 | 3.07(1) | 5+-1 |
|  |  |  |  |  |  | Ni-Me | 2.0 | 3.21(3) | 5+-1 |

Table 2 Distances Me-O and Me-Me.

|       | Me1(4g)    | Me2(2a)    | Me3(4h)    | Me4(2d)    |
|-------|------------|------------|------------|------------|
| Me-O  | 1,94       | 2,023 (2)  | 2,105      | 2,09 (2)   |
|       | 2,026      | 2,0877 (4) | 2,111      | 2,07 (4)   |
|       | 2,0979 (2) |            | 1,96 (2)   |            |
|       | 2,1115 (2) |            | 2, 06 (2)  |            |
| Me-Me | 2,9978 (2) | 2,9978 (2) | 2,77 (1)   | 2,77 (2)   |
|       | 3,0921 (2) | 3,02 (4)   | 2,9978 (2) | 2,9978 (2) |
|       | 3,11 (2)   |            | 3,02 (2)   | 3,09 (4)   |
|       | 3, 34 (2)  |            | 3,11 (2)   |            |
|       |            |            | 3,34 (2)   |            |

Table 3 Calculated values of superexchange integrals.

| ij | Angles | Expression | J |
|---|---|---|---|
| 4h-4h | $\alpha = 93°$ $\beta = 99°$ | $J = \dfrac{8}{3} bc J_{Mn}(\sin\alpha + \sin\beta)$ | -1,815 |
| 4g-4g | $\alpha = 90{,}4°$ $\beta = 91°$ | $J = \dfrac{8}{3} bc J_{Ni}(\sin\alpha + \sin\beta)$ | 5,570 |
| 2d-2d | $\alpha = \beta = 92°$ | | 5,564 |
| 2a-2a | $\alpha = \beta = 91°$ | | 5,568 |
| 4h-2d | $\alpha = 84°$ | $J = 2c\left(\left(c + \dfrac{1}{3}b\right)J_{Ni} - \dfrac{4}{3}b(U_{Ni} + U_{Mn})\right)\sin\alpha$ | -5,139 |
| 4g-4h | $\alpha = 95°$ $\beta = 99°$ | $J = \left[\left(c + \dfrac{4}{3}b\right)J_{Ni} - \dfrac{4}{3}b(U_{Ni} + U_{Mn})\right]c\sin\alpha +$ $+ \left[\left(c + \dfrac{1}{3}b\right)J_{Ni} - \dfrac{4}{3}b(U_{Ni} + U_{Mn})\right]c\sin\beta$ | -4,599 |
| 4h-2a | $\alpha = 92°$ $\beta = 98°$ | $J = c\left[\left(\dfrac{1}{3}b + c\right)J_{Ni} - \dfrac{4}{3}b(U_{Ni} + U_{Mn})\right](\sin\alpha + \sin\beta)$ | -4,618 |
| 4g-4h | $\alpha = 117°$ | $J = \left[\left(\dfrac{2}{3}b + c\right)J_{Ni} - \dfrac{2}{3}b(U_{Ni} + U_{Mn})\right]\sin\alpha +$ $+ \left[c^2 J_{Ni} + \dfrac{2}{3}b^2\left(J_{Mn} - \dfrac{1}{3}(U_{Ni} + U_{Mn})\right)\right]\cos\alpha$ | -1,010 |
| 4g-2a | $\alpha = 121°$ | $J = \dfrac{4}{3}b\left(cJ_{Ni}\sin\alpha - \dfrac{4}{3}b^2 U_{Ni}\cos\alpha\right)$ | -1,794 |
| 4h-2d | $\alpha = 165°$ | $J = 2\left[\dfrac{2}{3}b^2\left(J_{Mn} - \dfrac{1}{3}(U_{Ni} + U_{Mn})\right) + c^2 J_{Ni}\right]\cos\alpha$ | 0,558 |